# An attempt to control a manmade nuclear fusion

Yuri Kornyushin

*Maître Jean Brunschvig Research Unit, Chalet Shalva, Randogne, CH-3975*

A detailed simple model is applied to study a high temperature hydrogen plasma ball. It is assumed that the ions and delocalized electrons are distributed randomly throughout the charged plasma ball (extra/missing charge is assumed to be found in a thin layer on the surface of a ball). The energy of the microscopic electrostatic field around the ions is taken into account and calculated. It is shown in the framework of the model that charged hydrogen plasma ball can be stable as a metastable state, when subjected to external (atmospheric) pressure. Equilibrium radius of a ball, the barrier and the enthalpy of the equilibrium ball are calculated. It looks like the charged plasma ball in a metastable equilibrium should be used to conduct controllable nuclear fusion. Changes in the electric charge can be used to control the volume of a plasma ball.

## 1. Introduction

A manmade nuclear fusion occurs usually in high temperature artificial plasma balls. The plasma to be used for a nuclear fusion is deuterium-tritium plasma. We shall refer to it further as hydrogen plasma. It may be kept either at atmospheric pressure (when the aim is to perform a controlled nuclear fusion) or in a high-pressure vessel (as in the H-bomb).

We shall model the conditions as adiabatic ones under given pressure. May be this approach is more suitable for the H-bomb, when the *instability is planned* and the processes are quick. Anyway we shall restrict our consideration here by given entropy and given pressure condition. In this case the thermodynamic potential, having a minimum in the state of equilibrium, is the enthalpy (it is a function of thermodynamic variables, entropy $S$ and pressure $P$: $H(S,P)$) [1].

A general plasma ball was considered in [2]. Hydrogen plasma has some specific features. So here we shall analyze this specific case. First of all let us calculate the electrostatic energy of a separate ion.

## 2. Electrostatic energy of a separate ion

Let us consider a plasma ball of a volume $V = 4\pi R^3/3$ ($R$ is the radius of a ball), consisting of $n$ ions and $zn$ delocalized electrons (for a hydrogen high temperature, $T$, plasma $z = 1$). We consider here the ions as point charges, and the delocalized electrons like a negatively charged gas. We assume that all the electrons are not bound. As the ionization energy of the hydrogen atom is 13.598 eV ($T = 157782$ K) [3], this means that the temperature is rather high, about or higher than a quarter of a million of Kelvin degrees. At such a high temperature the plasma is classical. Classical delocalized electrons screen long-range electrostatic field of point charges. The screening Debye-Huckel radius is as follows [4]:

$$1/g = (kTV/4\pi zne^2)^{1/2} = 0.577 R^{3/2}(kT/zn)^{1/2}/e = R^{3/2}/R_0^{1/2}, \qquad (1)$$

where $e > 0$ is the elementary charge and $R_0 = 3zne^2/kT$.

The electrostatic field around a separate positive ion, submerged into the gas of classical electrons, is as follows [4]:

$$\varphi = (ze/r)\exp{-gr}, \qquad (2)$$

where $r$ is the distance from the center of the ion.

The electrostatic energy of this field is the integral over the ball volume of its gradient in a second power, divided by $8\pi$. The lower limit of the integral on $r$ should be taken as $r_0$, a very small value. Otherwise the integral diverges. Calculation yields the following expression for the electrostatic energy of a separate ion:

$$U_0 = 0.5z^2e^2(r_0^{-1} + 0.5g)\exp{-2gr_0}. \tag{3}$$

For $2gr_0$ essentially smaller than unity Eq. (3) yields:

$$U_0 = (z^2e^2/2r_0) - 0.75(z^2e^2g), \quad g = R_0^{1/2}/R^{3/2}. \tag{4}$$

Eq. (4) corresponds to Eq. (3) in [2]. Unfortunately, a factor of 0.75 in the right-hand part of Eq. (3) in [2] is missing.

The first term in the right-hand part of Eq. (4), $z^2e^2/2r_0$, represents the electrostatic energy of the bare ion. The hydrogen atom contains one proton only ($z = 1$). It should be noted that this term does not exist in the case of a hydrogen, deuterium or tritium atom, as it represents a self-action, which is zero in quantum mechanics, as is well known [5].

It is worthwhile to note that the expansion of a ball leads to the decrease in the delocalized electron density. From this follows the increase in the screening radius [see Eq. (1)]. The electrostatic energy of a separate ion increases concomitantly. One can see it, regarding Eq. (4).

### 3. Enthalpy of a charged plasma ball

We regard the ions of the considered plasma ball as randomly distributed. It is well known since 1967, that the electrostatic energy of $n$ randomly distributed ions is just $U = nU_0$ [6].

Let the extra charge of a ball be $eN$ (here $N$ is the positive/negative number of a missing/extra electrons). This charge is assumed to be situated on the surface of a ball (as the charges of the same sign repel each other). It produces electrostatic field outside the ball, $\varphi_e = (eN/r)$ [7]. Inside the ball this charge produces no field [7]. The electrostatic energy of the field of this extra/missing electrons charge is as follows [5,7]:

$$U_e = e^2N^2/2R. \tag{5}$$

The enthalpy of the whole system is as follows:

$$H(R) = (4\pi/3)PR^3 + (e^2N^2/2R) + (z^2e^2n/2r_0) - 0.75(z^2e^2nR_0^{1/2}/R^{3/2}), \tag{6}$$

where $P$ is the external (atmospheric) pressure. Again, the term $(z^2e^2n/2r_0)$ should be considered equal to zero for hydrogen plasma.

For a neutral plasma ball, when $N = 0$, $H(R)$ is a monotonic function of $R$. It increases monotonically with the increase in $R$. It has neither a maximum nor a minimum. So, the neutral plasma ball is not stable.

For a charged plasma ball when $P = 0$, $H(R)$ as a function of $R$ has a maximum only. One can see it, regarding Eq. (6). So at zero external pressure the plasma ball is not stable either.

At non-zero pressure $P$ the enthalpy $H(R)$ increases with the increase in $R$, may reach a maximum at some $R = R_{max}$, then it may reach a minimum at some $R = R_{min}$ (for a certain range of parameters), and then increases indefinitely. So, electric charge together with external pressure can stabilize the plasma ball size, when the parameters are suitable. Minimum of $H(R)$ corresponds to a metastable equilibrium. At $P = 0$ the enthalpy $H(R)$ never has a minimum, it has a maximum only.

At $R \leq R_{min}$ the system collapses, its radius goes smaller and smaller as a result of instability. May be this is what happens in the H-bomb.

Eq. (5) yields the following equation for the extremal values of $R$, $R_e$:

$$(\partial H/\partial R)_{R = Re} = 4\pi P R_e^2 - (e^2 N^2 / 2R_e^2) + 1.125(z^2 e^2 n R_0^{1/2}/R_e^{5/2}) = 0. \quad (7)$$

Let us consider now the ball in a vacuum ($P = 0$). At $P = 0$ Eq. (7) yields:

$$R_{max} = 5.0625 z^4 n^2 R_0 / N^4. \quad (8)$$

For $z = 1$, $n = 10^{21}$, $N = 10^{14}$, $kT = 6.9 \times 10^{-11}$ erg ($T = 500466$ K) we have $R_0 = 10^{13}$ cm and $R_{max} = 0.50625$ cm. This value of $R_{max}$ corresponds to the maximum of the enthalpy of a system. The density of the delocalized electrons at that is $N/V = 1.84 \times 10^{21}$ cm$^{-3}$. It is rather a high density. The enthalpy of the ball in a position of a maximum of $H(R)$ for selected values of parameters is $H(R_{max}) = 212.4$ J. It is not so high.

For a charged hydrogen plasma ball, subjected by atmospheric pressure ($P = 1$ bar), with $N = 10^{16}$, $n = 1.246 \times 10^{21}$, $z = 1$, and $T = 5 \times 10^5$ K we have $R_{max} = 7.53$ cm, $R_{min} = 25.4$ cm, $H(R_{max}) = 51.52$ kJ, $H(R_{min}) = 35.81$ kJ. This enthalpy is released in a case of disintegration of a charged hydrogen plasma ball. The barrier between the metastable equilibrium and unstable equilibrium [maximum of $H(R)$] is $H_{max} - H_{min} = 15.71$ kJ.

## 4. Discussion

When there is no external pressure, increase in the value of $R$ leads to the decrease in the electrostatic energy of a charge of a ball and to the increase in the electrostatic energy of the ions. This competition could be expected to yield the equilibrium value of a radius, $R_e$. But it is not so, this competition yields only a maximum of $H(R)$.

When external (atmospheric) pressure is not zero, it seems that the presence of the $4\pi P R_e^2$ term in Eq. (7) can yield the value of the equilibrium radius (additional work is required to expand the ball). It happens, but not at all the values of parameters.

It looks like it is more convenient to conduct a nuclear fusion reaction when the hydrogen plasma ball is charged and it has an equilibrium size (at non-zero external pressure). For that the plasma density should be high enough. Increase in the electric charge of a plasma ball changes the size of the ball and influences the nuclear fusion rate.

When the radius of a hydrogen plasma ball is smaller than $R_{max}$, the ball collapses and it is impossible to control the nuclear fusion rate.